\documentclass[prl,showpacs,byrevtex]{revtex4}
\input epsf
\usepackage{graphicx}
\usepackage{bm}

\def\dsodt{ds_{23}\over dt}
\def\dstdt{ds_{13}\over dt}
\def\dsthdt{ds_{12}\over dt}
\def\uoo{c_{13}c_{12}}
\def\uot{c_{13}s_{12}}
\def\uoth{s_{13}}
\def\uto{-c_{23}s_{12}-c_{12}s_{13}s_{23}}
\def\utt{c_{12}c_{23}-s_{12}s_{13}s_{23}}
\def\utth{c_{13}s_{23}}
\def\utho{s_{12}s_{23}-c_{12}s_{13}c_{23}}
\def\utht{-c_{12}s_{23}-c_{23}s_{13}s_{12}}
\def\uthth{c_{13}c_{23}}
\def\be{\begin{equation}}
\def\ee{\end{equation}}
\def\ba{\begin{eqnarray}}
\def\ea{\end{eqnarray}}
\def\gu{g_u}
\def\gd{ g_d}
\def\gup{\tilde g_u}
\def\gdp{\tilde g_d}
\def\br{\begin{array}}
\def\er{\end{array}}

\def\Dmott{\Delta m^2_{21}}
\def\Dmttht{\Delta m^2_{32}}

\begin{document}
\title{  Radiative Magnification of Neutrino Mixings in Split 
Supersymmetry}
\author{R. N. Mohapatra}
\email{rmohapat@physics.umd.edu}
\affiliation{Department of Physics and Center for String and Particle 
Theory, University of Maryland,
College Park, MD 20742, USA.}
\author{M. K. Parida}
\email{mparida@sancharnet.in}
\affiliation{Department of Physics, North Eastern Hill University,
Shillong 793022, India.}
\author{G. Rajasekaran}
\email{graj@imsc.ernet.in}
\affiliation{Institute of Mathematical Sciences, Chennai 600 113, India.}
\begin{abstract}
Radiative corrections to neutrino mixings in seesaw models depend
on the nature of new physics between the weak and the GUT-seesaw
scales and can be taken into account using the renormalization group 
equations. This new physics effect becomes
particularly important for models with quasi-degenerate neutrino
masses where small neutrino mixings at the seesaw scale can get
magnified by radiative renormalization effects alone to match 
observations. This mechanism of radiative magnification which
provides a simple understanding of why lepton mixings are so
different from quark mixings was demonstrated by us for the standard 
supersymmetry scenario where the
particle spectrum becomes supersymmetric above the weak scale. In
this paper, we examine this phenomenon in
split supersymmetry scenarios and find that the mechanism works also for 
this scenario provided the SUSY scale is at least 2-3 orders below the 
GUT-seesaw scale and one has larger values of ${\tan\beta}$.
\end{abstract}
\date{\today}
\rightline{UMD-PP-05-044}
\rightline{NEHU/PHY-MP-02/05}
\rightline{IMSc/2005/04/09}
\rightline{hep-ph/0504236}
\maketitle
\section{I. INTRODUCTION}
\label{sec1}
An important question in particle physics is the nature of new physics 
beyond the astoundingly successful standard model. One compelling 
scenario for this new 
physics is the weak scale supersymmetry which provides an answer to a 
number of puzzles of the standard model such as a solution to the gauge 
hierarchy problem, the origin of electroweak symmetry breaking as well as 
providing a candidate for dark matter of the universe. An added
virtue is that it unifies the disparate strengths of the weak,
electromagnetic and strong forces at a scale of about $10^{16}$ GeV, 
opening up another rich landscape of new physics around this 
scale. This 
high scale new physics can only manifest itself at low energies via large 
radiative 
correction effects which depend logarithmically on mass as well as 
through effects 
such as proton decay which are suppressed by this new high 
scale. 

There is reason to believe that there might be a 
manifestation of this new physics effect in the domain of neutrinos.
This arises from the consideration of seesaw mechanism\cite{seesaw} which 
provides a 
very natural way to understand the 
extreme smallness of neutrino masses and which requires the existence of 
massive 
right handed neutrinos with at least one having mass close to the GUT 
scale. It is therefore 
suggestive that both the seesaw scale and the GUT scale are one and the 
same. This connection becomes particularly plausible in the context 
of models 
based on the grand unifying groups such as  SO(10)\cite{so10} or 
$SU(2)_L\times 
SU(2)_R\times SU(4)_c$\cite{ps} that predict the existence of the right 
handed 
neutrino as part of the 
common fermion multiplet along with the fermions of the standard model.

In a recent paper, this concept of unification of GUT scale and seesaw 
scale was taken one step further by making the plausible hypothesis
that quark and lepton mixings angles may be same at the seesaw scale due 
to quark lepton unification whereas the observed large mixings at the 
weak scale are a consequence of radiative corrections. Using the formulae 
for renormalization group evolution for neutrino mass\cite{ref4,ref5}, it 
was shown that this possibility can be realized if the neutrino 
masses are quasi-degenerate and have same CP\cite{mpr1,mpr2}. Models based 
on $SU(2)_L\times SU(2)_R\times SU(4)_c$ group were presented where both the 
quasidegeneracy as well as mixing unification at the seesaw scale arose 
from the so-called 
type II seesaw formula\cite{seesaw2}. Detailed renormalization group 
evolution of mixing angles then revealed that one indeed gets the 
desired bilarge mixing pattern at low energies which are in agreement with 
the solar and the atmospheric neutrino data.
The third angle also undergoes radiative magnification, but remains small and
well within the CHOOZ-Palo-Verde limit\cite{ref3} due to the smallness of 
the corresponding initial value which is  the quark mixing,
$\sin\theta^0_{13}=0.0025$.

\par

Furthermore, our hypothesis has the interesting prediction that the 
common value for the quasi-degenerate neutrino mass is larger than 0.2 
eV, which can be tested in the next round of the neutrinoless double 
beta decay experiments\cite{bb} and is in the range claimed in 
Ref.\cite{ref9}. It also overlaps with the range accessible to the KATRIN 
experiment\cite{ref10} and consistent with the bound obtained from 
WMAP\cite{ref11}.
Threshold corrections at the low-energy SUSY scale have been found to 
improve agreement\cite{ref12} with the most recent data including KamLAND 
and SNO\cite{ref6} for a particular choice of the SUSY particle spectrum.

As mentioned, two important ingredients that allow the radiative 
magnification to work are low-energy 
supersymmetry(SUSY) and large values of $\tan\beta$  which contribute  
large logarithmic factors for radiative magnification via amplified 
values of $\tau$-Yukawa coupling.

Recently, a new unification scenario has been 
advocated\cite{ref13} which is based on the observation that one can 
maintain gauge-coupling unification and neutralino dark matter without 
necessarily 
``buying'' the entire machinary of weak scale supersymmetry but rather 
keeping only a subset of SUSY particles i.e. the gauginos and the Higgsino 
at the weak scale and pushing the rest of the superpartners to 
high intermediate scales($M_{\rm S}$). This has been called split 
supersummetry. This 
approach discards the naturalness requirement for the Higgs mass that 
motivated the weak scale supersymmetry- thus one has to fine tune the 
Higgs mass to every loop order. One also has no simple mechanism to 
understand the electroweak symmetry breaking. Despite these disadvantages,
one might consider this as an interesting minimal extension of the 
standard model that 
preserves gauge-coupling unification and provides a neutralino dark 
matter and discuss how neutrinos fit into it.

 Clearly, there is no 
inherent obstacle to implementing the seesaw mechanism and quark-lepton 
unification in these models. Only thing one has to keep in mind is that  
since the seesaw physics and quark-lepton unification generally tend to 
add new 
contributions to the gauge coupling evolution, if we want to retain 
the simple unification of three standard model gauge couplings,  
the seesaw scale must be at or above the GUT scale. In this 
paper, we work within a framework of this type and see if we can 
 understand the large neutrino mixings via the radiative 
magnification mechanism.

Below the high SUSY scale($M_{\rm S}$) in split supersymmetry, 
apart from the presence of gauginos and Higgsinos, the effective theory
is governed by nonSUSY Standard Model with one light Higgs doublet.
As the  two important ingredients in the radiative magnification 
scenario i.e. the long intervals of running in the 
presence of weak scale SUSY and 
 the enhancement factor due to $\tan\beta$ are missing in the split 
supersymmetry  below $M_S$, a natural apprehension emerges about 
the validity of 
this mechanism\cite{mpr1,mpr2} for large neutrino mixings.
Working within a model with quark-lepton 
unification based on the group $SU(2)_L\times SU(2)_R \times SU(4)_c$ 
 and degenerate masses for neutrinos obtained via the type II seesaw 
mechanism\cite{seesaw2, ref17} and the usual split SUSY spectrum, we show that this model does 
indeed lead to bilarge mixing pattern at low energies. The $SU(2)_L\times 
SU(2)_R \times SU(4)_c$ symmetry guarantees mixing unification between 
the quark and the lepton sectors at the seesaw scale (i.e. 
$\theta_{ij}^q(M_R)~=~\theta^\ell_{ij}(M_R)$).

 The new point which is 
different from the work of Ref.\cite{mpr2} is that running of neutrino 
masses is now different. Furthermore, in order to maintain gauge coupling 
unification, we must have the seesaw scale equal to the GUT scale, as 
already noted. We carry out this running effect and show that
despite the high split SUSY scale, the radiative magnification mechanism 
operates successfully to give the bilarge mixings at the weak 
scale provided that,
in addition to the quasi-degenerate Majorana neutrino masses with 
same CP, we have  the  scalar SUSY partners with masses starting
at least 2-3 orders  below the seesaw scale. Whereas with low-energy
SUSY, the radiative magnification occurs at $\mu \le 1$ TeV, in 
split supersymmetry it occurs at the high value of the SUSY scale which 
may be any where between $10^5$ GeV-$10^{15}$ GeV. Below this scale, the 
RGE running does not affect the neutrino mixings significantly. What is 
interesting is that radiative magnification occurs for the scalar 
susy partner masses close to even the scale $10^{15}$ GeV so that 
a major part of the running is nonsupersymmetric. Including small 
threshold effects 
needed to fit the details like the solar mass-squared difference, the 
 extrapolated low enery values of $\Dmttht$ , $\Dmott$ , and the mixing 
 angles are in excellent agreement with the current solar , atmospheric, 
and the reactor neutrino data.

This paper is organized as follows: in sec. 2  we discuss the relevant 
RGEs in the context of  split supersymmetry. In sec. 3,
we present allowed perturbative region of $\tan\beta$, radiative  
magnification of mixing
angles at high SUSY scales and  low energy extrapolation of masses and mixings.
In sec. 4, we discuss threshold effects and their estimations. Discussion 
of the results and conclusions are presented in sec.5.

\newpage

\par\noindent\\

\section{II. RENORMALIZATION GROUP EQUATIONS IN SPLIT SUPERSYMMETRY}
\label{sec2}

In this section while discussing renormalization group equations(RGEs) for 
neutrino masses and mixings
we point out some special features of split supersymmetry that contribute
to radiative magnification of the mixings at high SUSY scales.
Since we will follow the bottom-up approach to determine initial conditions at
the seesaw scales for RG-evolutions of neutrino parameters in the top-down
approach, we use the stadard RGEs for
the SUSY, split supersymmetry, and nonSUSY cases as applicable in appropriate
domains\cite{ref13,others,ref16}.
In the context of split supersymmetry where the scale($M_S$) of the scalar 
superpartner masses 
could be much larger than the
weak scale\cite{ref13}, the seesaw scale($M_R$) is clearly always larger 
than the SUSY scale($M_S$) i,e $M_R >> M_S$. We  assume  the
neutrino mass eigenstates to be quasi-degenerate and possess the same CP.
We also  ignore all CP
violating phases in the mixing matrix, and  adopt the mass ordering to be of
type $m_3 \agt m_2 \agt
 m_1$. Parametrizing  the  $3\times 3$ mixing matrix as

\par
\noindent
\be U=\left[\br{ccc}
\uoo&\uot&\uoth\\
\uto&\utt&\utth\\
\utho&\utht&\uthth
\er\right],\label{eq1}\ee

\par
\noindent
where $c_{ij}=\cos\theta_{ij}$ and $s_{ij}=\sin\theta_{ij} (i,j=1, 2,
3$), the RGEs for the mass eigen values and mixing angles  can be written
as \cite{ref4,ref5}

\par
\noindent
\be{dm_i\over dt}=-2F_{\tau}m_iU_{\tau
i}^2-m_iF_u = b^{(m)}_im_i,\,\left(i=1,2,3\right).\label{eq2}\ee

\par
\noindent
\ba\dsodt&=&-F_{\tau}{c_{23}}^2\left(
-s_{12}U_{\tau1}D_{31}+c_{12}U_{\tau2}D_{32}
\right),\label{eq3}\\
\dstdt&=&-F_{\tau}c_{23}{c_{13}}^2\left(
c_{12}U_{\tau1}D_{31}+s_{12}U_{\tau2}D_{32}
\right),\label{eq4}\\
\dsthdt&=&-F_{\tau}c_{12}\left(c_{23}s_{13}s_{12}U_{\tau1}
D_{31}-c_{23}s_{13}c_{12}U_{\tau2}D_{32}\right.\nonumber \\
&&\left.+U_{\tau1}U_{\tau2}D_{21}\right).\label{eq5}\ea
\par
\noindent
Here $D_{ij}={\left(m_i+m_j)\right)/\left(m_i-m_j\right)}$ and,~for MSSM with  
$\mu \ge M_{\rm S}$,

\par

\noindent

\ba F_{\tau}&=&{-h_\tau^2}/{\left(16\pi^2\cos^2\beta\right)},\nonumber\\
F_u&=&\left(1\over{16\pi^2}\right)\left({6\over5}g_1^2+6g_2^2-
6{h_t^2\over\sin^2\beta}\right),\label{eq6}\ea
\par
\noindent
but, for $\mu \le M_{\rm S}$,

\par
\noindent
\ba F_{\tau}&=&{3h_\tau^2}/\left( 32\pi^2\right),\nonumber \\
F_u&=&\left(3g_2^2-2\lambda-6h_t^2-2T \right)/\left(16\pi^2\right).
\label{eq7}\ea

\par
\noindent
with the definitions of the couplings at $M_{\rm S}$,
\par
\noindent

\ba \gu\left(M_{\rm S}\right)&=&g_2\left(M_{\rm S}\right)\sin\beta,
~~\gd\left(M_{\rm S}\right)=g_2\left(M_{\rm S}\right)\cos\beta,\nonumber\\
 \gup\left(M_{\rm S}\right)&=&\left(3/5\right)^{1/2}g_1\left(M_{\rm S}\right)
 \sin\beta,~~\gdp\left(M_{\rm 
S} \right)=\left(3/5\right)^{1/2}g_1\left(M_{\rm 
S}\right)\cos\beta.\label{eq8}\ea

\par
\noindent
The  additional term present in eq.(7) which is specific to the 
split SUSY below $M_{\rm S}$ and the Higgs quartic coupling  are,
\par
\noindent
\ba  T&=&{3\over2}\left({\gu}^2+{\gd}^2\right)+ 
{1\over2}\left({\gup}^2+{\gdp}^2\right),\nonumber\\
\lambda(M_{\rm S})&=&{\left[g_2^2(M_{\rm S})+(3/5)g_1^2(M_{\rm 
S})\right]\over4}\cos^2 2\beta.\label{eq9}\ea

\newpage

\par
\noindent\\

The basic mechanism  responsible for  radiative magnification of mixings of
quasi-degenerate neutrinos with same CP in MSSM  
which has been pointed out earlier\cite{mpr1,mpr2} is also applicable 
with split SUSY but with  high value of $M_{\rm S}$.
Since $m_i$ and $m_j$ are scale dependent, the
initial difference existing between them at $\mu=M_R$ is narrowed down
during
the course of RG evolution as we approach the SUSY scale. This causes
$D_{ij}\to$ large, and hence large magnification to the mixing angle due to
radiative effects through eqs.(3)-(5). Another major factor contributing to radiative
magnification  is due to amplified negative value  of $F_\tau$ in eqs.(3)-(5)
in the presence of SUSY by a factor $\simeq \tan^2\beta$
in the large $\tan\beta$ region.

An approximate estimation of values of $\tan\beta$ needed for radative magnification in split SUSY is possible  by noting that the magnification has been 
realized in MSSM with weak-scale SUSY for  $\tan\beta \approx 50-55$
and in that case the  major  factor
controlling magnification
is due to amplification of $\tau$-Yukawa coupling\cite{mpr1,mpr2}. 
Expecting similar amplification in split SUSY  gives the appoximate relation,
  
\par
\noindent
\ba {\left(h^0_{\tau} \tan\beta^0\right)}^2 \log(M^0_R/M^0_S)&=&
    {\left( h_{\tau}\tan\beta\right)}^2 \log(M_R/M_S). \label{eq10}\ea
\par
\noindent
where quantities with zero superscript(without superscript) refer to
weak-scale(split) supersymmetry. Noting that the ratio of the two $\tau$-Yukawa
coplings in eq.(10) is $\approx O(1)$, and using $\tan\beta^0= 50-55$\cite{mpr1, mpr2}, and $M^0_R=
M_R=M_U=2\times 10^{16}$ GeV, the above relation gives the approximate
requirements for split supersymmetry case as $\tan\beta=75$, and $110$ for 
$M_{\rm S}=10^9$ GeV, and $10^{13}$ GeV, respectively. In fact, it has been
noted that  split SUSY permits larger values 
of $\tan\beta$ than the MSSM with weak scale SUSY\cite{drees}.   

It would be shown in Sec.3 that the allowed perturbative upper limit of 
$\tan\beta$ increases with increase of the SUSY scale.
Thus even though the supersymmetric part of the running of neutrino
parameters is much less than in the weak scale supersymmetry case,
larger values of tan beta increases the value of $F_\tau$ and this in
turn leads to the desired enhancement of the radiative magnification
effect on the mixing angles.


 In addition to the above, here we
point out that there is a specific mechanism which operates  in the 
allowed parameter space of
split SUSY that causes  mass eigenvalues to approach one 
another faster and thus,
drives radiative magnification even if the 
SUSY scale is many orders larger than the weak scale.
Noting that the $\beta$-function coefficient for the evolution of mass 
eigenvalue $m_i(\mu)$ in eq.(2) is $b^{(m)}_i$=
$ -2F_{\tau}U_{\tau i}^2-F_u$, near the seesaw scale $U_{\tau i}^2\sim 0$,
for $i=1,2$ because of high-scale mixing unification constraint, but
$U_{\tau3}^2\sim 1$. Therefore, the evolution of the third mass eigen value would be different from the first two. Further, since supersymmetric $F_{\tau}$
is negative for $\mu > M_{\rm S}$ and,
depending upon the allowed values of the gauge and the Yukawa coupling 
constants contributing to $F_u$ in eq.(6)
above the high SUSY scale, the $\beta$-function coefficient may be positive or negative thus resulting in  the  decreasing or 
increasing 
behavior of the mass-eigen value below the seesaw scale down to $M_{\rm S}$.
In particular with split supersymmetry $F_{\rm u}$ may be positive 
due to dominance of gauge couplings over the top-Yukawa coupling  
resulting in negative values of $b^{(m)}_{1,2}$ for $\mu > M_{\rm S}$.
But since  $F_\tau$ is negative, the positive value of the  first term,  
$ -2F_{\tau}U_{\tau i}^2$, may partly or largely cancel with the 
second term leading to a large or small positive value of $b^{(m)}_3$.  
As a result, the first two mass eigen values may be expected to increase but 
the ~third may decrease, rapidly or slowly, from their input values  at
 $M_{\rm R}$. 
This increasing behavior of some mass eigen values($m_1, m_2$) in the region
$M_{\rm S} < \mu < M_{\rm R}$  would be in sharp contrast to that in MSSM
with weak-scale SUSY where the allowed high scale values of the gauge and 
the Yukawa couplings and the mixing unification constraint have been found to 
make all the three $\beta$-function 
coefficients positive resulting in the decrasing behavior of the three mass 
eigen values below the seesaw scale although with different rates\cite{mpr2}.
Thus, with split SUSY the allowed values of the parameter space
may cause three mass eigen values  
to approach sufficiently closer to  one another even after much shorter 
interval of running. This may occur at  high values of 
$M_{\rm S}$  which may even be only $2-3$ orders smaller than $M_{\rm R}$.
Then the functions $D_{ij}\to$ large near $M_{\rm S}$ causing
magnification  of the mixing angles.

 On the otherhand for $\mu < M_{\rm S}$, the $\tau$-Yukawa coupling has 
negligible contribution and  the $\beta$-function coefficients are nearly 
the same for all
the three eigen values and
approximately equal to  $-2F_u$. From eq.(7)  is clear  that 
in the presence of nonSUSY SM, $F_u$
is negative below  $M_{\rm S}$ resulting in positive beta-function 
coefficients for 
all mass-eigen values causing them to decrease approximately in the same manner~down to  $M_{\rm Z}$. It is easy to check, using boundary conditions given in
eqs.(8)-(9), that  $F_u$ is negative at the starting point $M_{\rm S}$
of the nonSUSY theory,
with 
$16\pi^2F_u(M_{\rm S})=-{3\over5}g_1^2(M_{\rm S})-2\lambda(M_{\rm S})- 
6h_t^2(M_{\rm S})$.                                 
Although this  feature is also common to  MSSM with 
the weak-scale SUSY, in  split SUSY the  running interval
for nonsupersymmetric theory  is much  larger.
 
Our numerical solutions to masses and mixing angles reported in Sec.3
with split SUSY are found to corroborate these properties of the 
RG-evolution.

\newpage

\par\noindent\\

\section{III. RADIATIVE MAGNIFICATION AT HIGH SUSY SCALES}
\label{sec3}
 In this section we discuss the realization of radiative magnification of
neutrino mixing angles at high SUSY scales and low energy extrapolations
of mass eigen values and mixing angles for comparison with the available 
experimental data.

\par\noindent{\bf {III.1. Perturbative limit on $\tan\beta$}}\\
The fact that grand unification with split supersymmetry allows
larger values of $\tan\beta$ than those with weak-scale SUSY has been 
also noted in ref.\cite{drees}. 
Since the value of $\tan\beta$ is an important ingredient, at first we
estimate the maximum  allowed values of $\tan\beta$ for every value of
$M_{\rm S}$=$10^{3}$ GeV - $10^{15}$ GeV. For this purpose we use the
 standard PDG values of masses and couplings\cite{PDG} at $\mu=M_{Z}$ 
and follow the 
bottom-up approach scanning the values of the three gauge couplings and 
the third generation Yukawa couplings while varying input values of 
$\tan\beta$ over a wide range
at all values of $\mu$ up to the GUT-scale using the RGEs
in the appropriate regions\cite{ref13,others,ref16}. The lowest allowed value 
of 
$\tan\beta$ is determined when the top-quark Yukawa coupling reaches the 
perturbative limit($h_t=3.54$). Similarly the upper limit is determined 
by noting that the
$\tau$-lepton Yukawa coupling attains its perturbative limit at the GUT-scale.
 These are shown in Fig.1. It is clear that the upper limit on
$\tan\beta$ is larger compared to the weak-scale SUSY model and 
this limit also  increases quite  significantly with increasing value 
of $M_{\rm S}$
in split supersymmetry. For examle wth $M_{\rm S}$=$10^7$ GeV, $10^{9}$ GeV,
$10^{13}$ GeV, and $10^{15}$ GeV, 
, the upper limits are found to be  $\tan\beta = 85, 100, 132$, and $160$, 
respectively.
Through this bottom-up approach we also noted the values of different 
coupling constants and CKM mixings at different values of the seesaw 
scales starting from their low-energy values. Some of the coupling 
constants at the seesaw scales are presented in Table.1 where the factors 
like $\sin\beta$ ~or $\cos\beta$ are included in the values of SUSY Yukawa 
couplings.
With large values of $\tan\beta$ allowed near the upper limit, the Higgs mass
prediction in split SUSY tend to be independent of this parameter
and the two Yukawa couplings, $g_d$ and $\tilde {g_d}$, have negligible
effects below $M_S$.\\

\par\noindent{\bf{III.2.Radiative Mgnification of Mixing Angles}}\\
The values of  gauge and Yukawa couplings
, the CKM mixing angles
obtained at the seesaw scale from the bottom-up approach and  finetuned
values of light neutrino masses($m_i^0, i=1,2,3$) are  used as inputs
to obtain solutions from the  neutrino RGEs  in the top-down approach.~These 
 input parameters are given in Table 1.
~The unknown
parameters $m^0_i$
are determined in such a way that the magnified values of mixing angles
are at first obtained at $M_{\rm S}$ along with the mass eigen-values. These 
are further
extrapolated down to $\mu=M_{Z}$ through the corresponding RGEs to give
closest agreement with the experimental data:

\par
\noindent
\ba
\Dmott&=&\left(5-8\right)\times 10^{-5}{\rm eV}^2,
    ~\Dmttht=\left(1.2-3\right )\times10^{-3}{\rm eV}^2,\nonumber \\
    \sin\theta_{23}&=&0.67-0.707,~ \sin\theta_{12}= 0.5-0.6,
    ~\sin\theta_{13}\le 0.16~.\label{eq11}\ea
\par

As noted in Sec.2 the
signs and values of the three $\beta$-functions for the mass
eigen values at the respective seesaw scales are easily checked
from the initial values of couplings given in Table 1 using eq.(2). For 
example for the case with $M_{\rm R}=
2\times 10^{16}$ GeV, $M_{\rm S}=10^9$ GeV, $\tan\beta=90$, the values of 
coupling constants given in Table.1 yield the  
 $\beta$-function coefficients at $M_R$ to be  $b^{(m)}_3=0.0016$, 
$b^{(m)}_1 = b^{(m)}_2 = -0.0115 $ which, as per the predictions
of Sec.2, are responsible for the  
slow decrease of $m_3$ and rather faster increase of $m_1$ and $m_2$ below 
the seesaw scale. These features are clearly displayed in  
Fig.2. Similarly for   $M_{\rm R}=
2\times 10^{16}$ GeV, $M_{\rm S}=10^{13}$ GeV,  $\tan\beta=90$,
  these coefficients are $b^{(m)}_3\simeq 0.0015$, 
$b^{(m)}_1 = b^{(m)}_2 = -0.01165 $ at the GUT-seesaw scale. 
 It is quite interesting to note that
the mutual differences are narrowed down by the increase of the first and 
the second mass eigenvalues  and 
simultaneous decrease of the third eigen value
leading to large $D_{ij}$ and  the magnification at $M_{\rm S}$.
This may be contrasted with the low-energy SUSY case where all the three 
masses showed decreasing behavior although with different rates\cite{mpr2}.
This increasing behavior of neutrino masses $m_{1,2}$
in contrast to the decrasing behavior in low-energy SUSY is due to
different relative values of the gauge and Yukawa couplings at higher 
scales in the large
$\tan\beta$-region in split SUSY some of which are given in Table.1.
Below $M_{\rm S}$ the decrease is almost uniform
due to approximate equality of the corresponding $\beta$-function
~coefficients of the three mass eigenvalues in this region
as discussed in Sec.2.
The solutions shown in the Fig.4 and Fig.5   clearly exhibit radiative
magnification to bilarge mixings at the high values of $M_{\rm S}$=$10^9$ GeV,
 and
$M_{\rm S}$=$10^{13}$ GeV corresponding to the evolutions of mass eigen values 
as shown in Fig.2, and Fig.3, respectively.\\

\par\noindent\\

For the seesaw scale at 
$M_{\rm R}=M_{\rm {Pl}}= 2\times 10^{18}$ GeV the magnification of mixings
takes place even at such high SUSY-scale as $M_{\rm S}=10^{15}$ GeV.

\par
It is clear that due to larger allowed values of $\tan\beta$ and also due to
opposite signs of $\beta$-function coefficients of mass eigen values at high 
scales, the radiative magnification of neutrino mixings is possible in split
SUSY at high SUSY scales  even if the SUSY-scale is onle $2-3$ orders smaller than the seesaw scale.   
 We find that although
enhancement due to RG evolution occurs in the $\nu_e-\nu_{\tau}$
sector also, the predicted low energy value remains at 
$\sin\theta_{13}=0.08-0.1$
which is well within the CHOOZ-Palo Verde bound\cite{ref3} and can be 
tested in the planned $\theta_{13}$ search experiments.

\section{IV. THRESHOLD EFFECTS }
\label{sec4}
In the previous section we considered 
the mass squared differences obtained at lower energies purely by 
RG-evolution and these, as ~shown in Table 1, are  in agreement
with gross features of the experimental data on atmospheric neutrinos but 
falls somewhat on the
higher side of
solar neutrino data. Since threshold effects have been shown to make 
significant
contribution on quasidegenerate neutrinos\cite{ref12,ref18,ref19}
in this section we
estimate them to spplement our RG-solutions of Sec.3.
 In particular we note that the  
loop facotr
at the  high-SUSY-scale threshold($M_{\rm, S}$), where the heavy superpartners 
are located, assumes a  very simple form. The threshold loop factor 
for neutrino mass at the electroweak scale is similar to  the 
SM\cite{ref19}  

In contrast to the weak-scale SUSY where the superpartners contribute 
predominantly  
near $M_{\rm Z}$\cite{ref18,ref19}, in split SUSY the
dominant threshold corrections to neutrino masses occur near high SUSY-scale 
threshold  $\mu_0 \sim M_{\rm S}$.
 Since the mixing angles in the 
radiative magnification scenario are very nearly the same  at both 
the thresholds,
it is quite convenient to use the approximation for the PMNS matix
$U_{\alpha{\rm i}}(M_{\rm S}) \sim  U_{\alpha{\rm i}}(M_{\rm Z})$
and evaluate them in the limit $\theta_{13} \to 0$ and $\theta_{23}  \to
 \pi/4$ which are very closely cosistent with our solutions\cite{ref12}.

Using the loop factors  $T_{ij}$ in the
mass basis with ($i,j=1,2,3$)  the threshold effects on mass eigen values
at the threshold $\mu_0$ are expressed as\cite{ref18,ref19},
\par
\noindent
\be
m_{ij}(\mu_0)=m_i(\mu_0)\delta_{ij}+m_i(\mu_0) T_{ij}(\mu_0)+m_j(\mu_0)
T_{ji}(\mu_0),(\mu_0= M_{\rm Z}, M_{\rm S}) \label{eq12}\ee
\par
\noindent
Transforming the loop factors  in terms of those in the
flavor basis,
$T_{\alpha\beta}(\alpha, \beta= e, \mu, \tau)$, it is straightforward
to evaluate the effects in the limiting case and for quasidegenerate neutrinos
with slightly different values of the common mass $m(\mu_0)$  at
$\mu_0= M_{\rm S}$, $M_{\rm Z}$ leading to the formula\cite{ref12}

\ba (\Delta m_{21}^2)_{th}(\mu_0)
=4\rm m^2(\mu_0)\cos 2\theta_{12}[-T_e(\mu_0) 
+(T_\mu(\mu_0)+T_\tau(\mu_0))/2], \label{eq13}\\
(\Delta m_{32}^2)_{th}(\mu_0)
=4\rm m^2(\mu_0)\sin^2\theta_{12}[-T_e(\mu_0) 
+(T_\mu(\mu_0)+T_\tau(\mu_0))/2], \label{eq14}\ea

\par
\noindent
When corrections at both the thresholds are included along with the 
contribution due to RG-evolution, the analytical formula for the mass 
squared differences
is expressed as,

\par
\noindent

\ba  (\Delta m_{ij}^2)( M_{\rm Z})= (\Delta m_{ij}^2)_{\rm RG}
 +(\Delta m_{ij}^2)_{\rm th}( M_{\rm Z})+(\Delta m_{ij}^2)_{\rm th}( M_{\rm S})
, (i>j = 1,2,3), \label{eq15}\ea

\par
\noindent
where the first term in the RHS of eq.(15) has been already estimated 
from RG-evolution and the second and the third terms are estimated 
through eqs.(13)-(14).
As the RG-evolution effects estimated in Sec.3 have already given approximately the correct values
of $\Dmttht$ for atmospheric neutrino data and $\Dmott$ only $4\sigma-5\sigma$
larger than the KamLAND and SNO data, small and simpler threshold effects
might be sufficient to account for these deviations. In 
split supersymmetry
such  contributions can easily arise from the high-scale SUSY-threshold 
effects at $\mu_0=M_{\rm S}$.

\par
Since $M_{\rm S}\gg M_{\rm Z}$ in split supersymmetry, left-handed charged
sleptons and sneutrinos have almost identical masses.
Denoting the masses of the left-handed  sleptons by 
$M_{\tilde \ell^L_{\alpha}}$
and the right-handed charged sleptons mass by   
$M_{\tilde \ell^R_{\alpha}}$ we obtain a simple formula for the loop
~factors  at the high SUSY scale($\mu_0=M_{\rm S}$),

\par
\noindent

\ba 16\pi^2 T_\alpha(M_S)&=&(3/8)(g^2_1(M_S)/5+g^2_2(M_S))
\left(ln(M_{\tilde \ell^L_{\alpha}}^2/M_S^2)-1/2\right) \nonumber\\
&&+\delta_{\alpha\tau}(1/4)h_\tau^2(M_S)(1+\tan^2\beta
 )\left(ln(M_{\tilde \ell^R_{\alpha}}^2/M_S^2)-
1/2\right).\label{eq16} \ea\\

\par\noindent\\

Ignoring the low-energy threshold effect at $M_{\rm Z}$  we find that 
with $M_{\tilde \ell_{\alpha}}$, or   $M_{\tilde \ell^R_{\alpha}}$ 
 few times lighter than $M_{\rm S}$
gives the desired threshold corrections with negative sign. For, example,
using a simple and plausible choice of approximate degeneracy in the
scalar superpartner spectrum around $M_{\rm S}$ with $M_{\tilde e_L}=
M_{\tilde \mu_L}=M_{\tilde \tau_L}=M_{\tilde e_R}=M_{\tilde \mu_R}$,
we have $[-T_e(M_S) 
+(T_\mu(M_S)+T_\tau(M_S))/2]\simeq (h_\tau^2(M_S)\tan^2\beta
 )[ln(M_{\tilde \tau_R}^2/M_S^2)-1/2]/(128\pi^2)$. Then with 
$M_{\tilde \tau_R}/M_S=0.5-1.1$, numerical values of  threshold corrections 
are obtained    
as shown in Table.1. Such threshold contributions, when added 
to the 
RG-effects, bring the theoretical predictions  in concordance with all
the available neutrino data including 
those from KamLAND and SNO.     

\par
\section{V. DISCUSSION AND CONCLUSION}
\label{sec5}

    In this section we discuss the implications of our results briefly and
 state our conclusions.
Our solutions require quasi-degenerate neutrino mass eigen values in the 
range $0.15 ~eV < m_i(M_Z) < 0.5 ~eV$ leading to the effective mass
in neutrinoless double beta decay given by    $|<M_{ee}>|$=$|\Sigma_i m_i 
U^2_{ei}|$= $0.15$ ~eV -$0.5$ ~eV.
This is accessible to all the experiments being planned to search for 
the neutrinoless double beta decay.
~Furthermore, this range of neutrino masses also
overlaps with the range accessible to the KATRIN\cite{ref10} 
 Tritium beta decay experiment.
The prediction $U_{e3}\equiv \sin\theta_{13}$=$0.08-0.10$
is also accessible to several planned long-baseline
neutrino experiments\cite{bnl} as well as the planned reactor 
experiments\cite{reactor}.
As discussed in ref.\cite{mpr2} the range of  eigenvalues of 
neutrino masses is  consistent with
WMAP observations  and also with the combined analysis of
WMAP$+$2dF GRS data\cite{ref11}.

\par\noindent\\

 In summary, we had shown previously that in the MSSM with weak-scale 
SUSY\cite{mpr2}, 
the hypothesis of quark-
lepton mixing unification and quasi-degenerate neutrino spectrum at   
the seesaw scale successfully explains
 the observed mixing pattern for neutrinos i.e. two large
mixings needed for $\nu_e-\nu_\mu$
and  $\nu_\mu-\nu_\tau$ and small mixing for $U_{e3}$ at low energies.
In this paper we have extended this discussion to the case of  split 
supersymmetry where we find that, despite the absence of low-energy SUSY 
and the corresponding absence of RG-running   with an amplified value 
of $\tau$-Yukawa coupling  over a considerable mass interval 
($M_Z$-$M_{\rm S}$), radiative magnification of neutrino mixings 
does occur. 
These deficits seem to be adequately compensated by
the larger values of $\tan\beta$ allowed in this case and the positive and 
negative values of $\beta$-function coefficients for the running of 
mass eigen values at high scales.  The key tests of the model still remain 
the common mass of neutrinos above $0.15$ eV  and $U_{e3}$ between 
0.08-0.1, as in \cite{mpr2}. 

\begin{table*}
\caption{Radiative magnification to bilarge neutrino mixings
 for input values of ${\rm m}^0_{\rm i}$($i=1,2,3$),
 $\sin\theta_{23}^0=0.038$,~$\sin\theta_{13}^0=0.0025$, ~and 
$\sin\theta_{12}^0=0.22$ at the high scale $M_{\rm R}$. The initial values
of SUSY Yukawa couplings at seesaw scales include factors $\sin\beta$ or 
$\cos\beta$ as applicable.} 
\begin{ruledtabular}
\begin{tabular}{lcccc}\hline
$M_{\rm R}$(GeV)&$2\times 10^{16}$&$2\times 10^{16}$&$2\times 10^{16}$&

$2\times10^{18}$\\

$M_{\rm S}$(GeV)&$10^{13}$&$10^{13}$&$10^9$&$10^{15}$\\

$\tan\beta$&$90$&$130$&$90$&$140$\\

$g_1^0$&$0.6206$&$0.6200$&$0.6518$&$0.6540$\\

$g_2^0$&$0.6203$&$0.6198$&$0.6522$&$0.6125$\\

$g_3^0$&$0.6262$&$0.6260$&$0.6565$&$0.5935$\\

$h_t^0$&$0.3943$&$0.4035$&$0.4450$&$0.3625$\\

$h_b^0$&$0.5325$&$0.9627$&$0.8052$&$0.9399$\\

$h_{\tau}^0$&$1.0181$&$2.2676$&$1.7592$&$2.5684$\\

$m_1^0$(eV)&$0.4483$&$0.2965$&$0.2267$&$0.3648$\\

$m_2^0$(eV)&$0.4500$&$0.30$&$0.2300$&$0.3700$\\

$m_3^0$(eV)&$0.4911$&$0.2965$&$0.3188$&$0.5060$\\

$m_1$(eV)&$0.2934$&$0.1938$&$0.1766$&$0.2301$\\

$m_2$(eV)&$0.2937$&$0.1944$&$0.1773$&$0.2310$\\

$m_3$(eV)&$0.2956$&$0.1983$&$0.1816$&$0.2364$\\

$(\Delta m^2_{21})_{\rm RG}$(eV$^2$)&$1.693\times 10^{-4}$&$2.28\times 10^{-4}$

&$2.31\times 10^{-4}$&$3.96\times 10^{-4}$\\

$(\Delta m^2_{32})_{\rm RG}$(eV$^2$)&$1.25\times 10^{-3}$&$1.53 \times 10^{-3}$

&$1.55\times 10^{-3}$&$2.53\times 10^{-3}$\\

$M_{\tilde \tau_R}/M_S$&$1.1$&$0.92$&$0.50$&$1.03$\\

$(\Delta m^2_{21})_{\rm th}$(eV$^2$)&$-0.84\times 10^{-4}$&$-1.48\times 10^{-4}

$&$-1.51\times 10^{-4}$&$-3.16\times 10^{-4}$\\

$(\Delta m^2_{32})_{\rm th}$(eV$^2$)&$-0.61\times 10^{-4}$&$-0.11\times 

10^{-3}$&$-0.10\times 10^{-3}$&$-0.19\times 10^{-3}$\\

$(\Delta m^2)_{\rm sol}$(eV$^2$)&$8.5\times 10^{-5}$&$8.0\times 10^{-5}$&

$8.0\times 10^{-5}$&$8.0\times 10^{-5}$\\

$(\Delta m^2)_{\rm atm}$(eV$^2$)&$1.19\times 10^{-3}$&$1.42\times 10^{-3}$&

$1.45\times 10^{-3}$&$2.34\times 10^{-3}$\\

$\sin\theta_{12}$&$0.545$&$0.5474$&$0.549$&$0.526$\\

$\sin\theta_{23}$&$0.701$&$0.703$&$0.707$&$0.707$\\

$\sin\theta_{13}$&$0.101$&$0.101$&$0.103$&$0.104$\\\hline
\end{tabular}
\end{ruledtabular}
\end{table*}
\label{tab1}

\par
\begin{figure}
\epsfxsize=8.5cm
\epsfbox{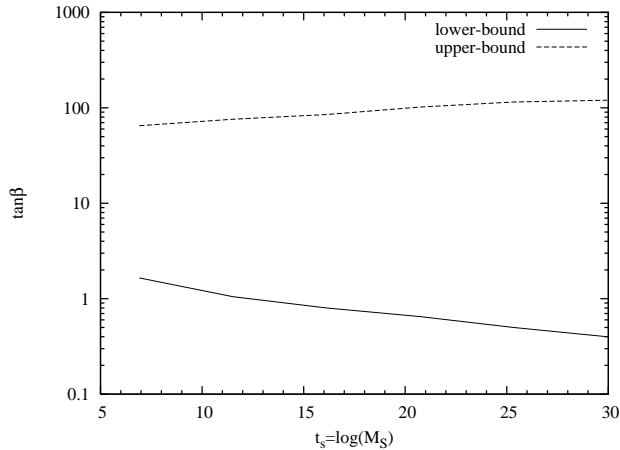}
\caption{Perturbative upper and lower limits on the allowed values of  
$\tan\beta$ as a function
of SUSY scale $M_{\rm S}$, taken in unit of GeV, in split supersymmetry.}
\label{fig1}
\end{figure}
\begin{figure}
\epsfxsize=8.5cm
\epsfbox{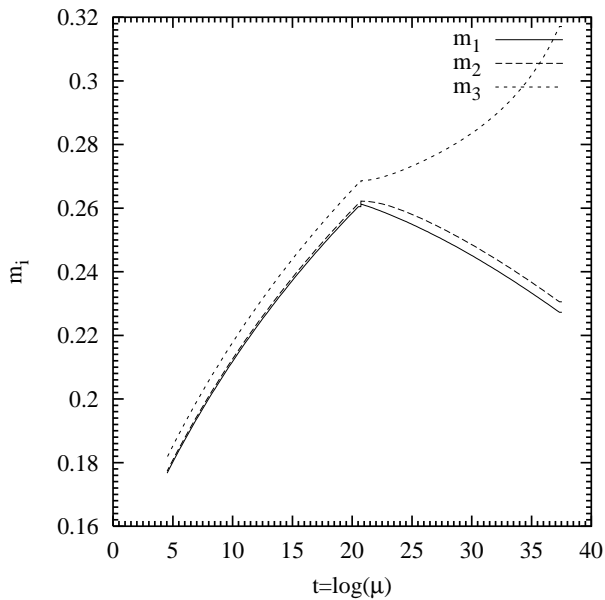}
\caption{Renormalization group evolution of light Majorana neutrino mass
eigenvalues
showing both the increasing and decreasing behaviors between SUSY scale
$M_{\rm S}=10^{9}$ GeV  and the GUT-seesaw scale $M_{\rm R}=2\times 10^{16}$ 
GeV  in 
split supersymmetry  where $t=\log(\mu)$ and $\mu$ is in
unit of GeV. The input values are given
in the fourth column of Table 1.}
\label{fig2}
\end{figure}
\begin{figure}
\epsfxsize=8.5cm
\epsfbox{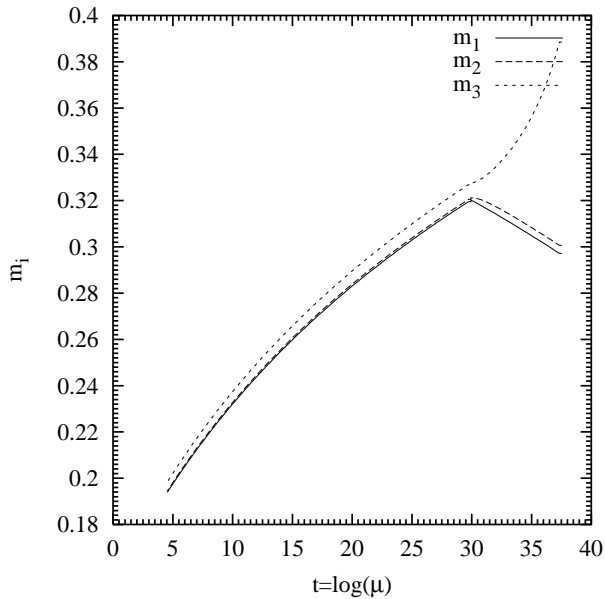}
\caption{Same as Fig.2 but with
$M_R=2 \times 10^{16}$ GeV and  SUSY scale $M_{\rm S}=10^{13}$ GeV.
 The input values and low energy extrapolations are given in the third column of Table 1.}
\label{fig3}
\end{figure}
\begin{figure}
\epsfxsize=8.5cm
\epsfbox{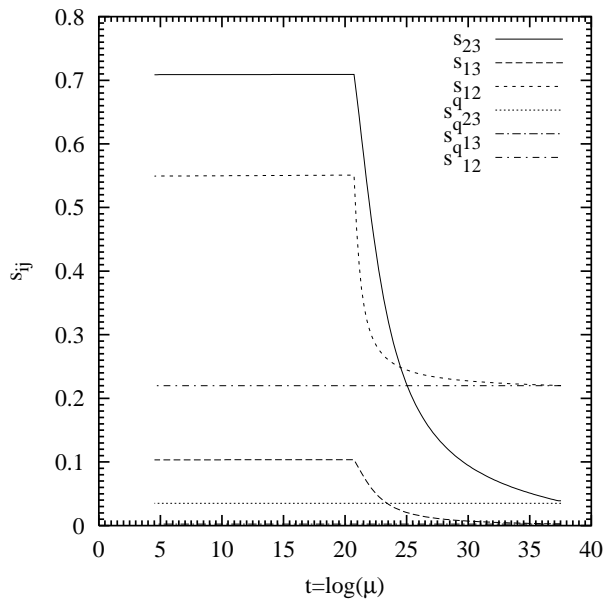}
\caption{Evolution of small quark-like mixings at the GUT-seesaw scale,
$M_R=2\times 10^{16}$ GeV to bilarge neutrino mixings at the SUSY scale
$M_S=10^{9}$ GeV, and extrapolation to low energies for the
input and output mass-eigen values and mixing angles given in the fourth column
~of Table 1. The solid, long-dashed, and short-dashed
lines represent the sines of neutrino mixing angles 
$\sin\theta_{23}$, $\sin\theta_{13}$, and $\sin\theta_{12}$, respectively,
as defined in the text. Almost horizontal lines originating from the seesaw
scale represent the sines of corresponding
 CKM mixing angles in the quark sector. The mass scale $\mu$ is in GeV}
\label{fig4}
\end{figure}
\begin{figure}
\epsfxsize=8.5cm
\epsfbox{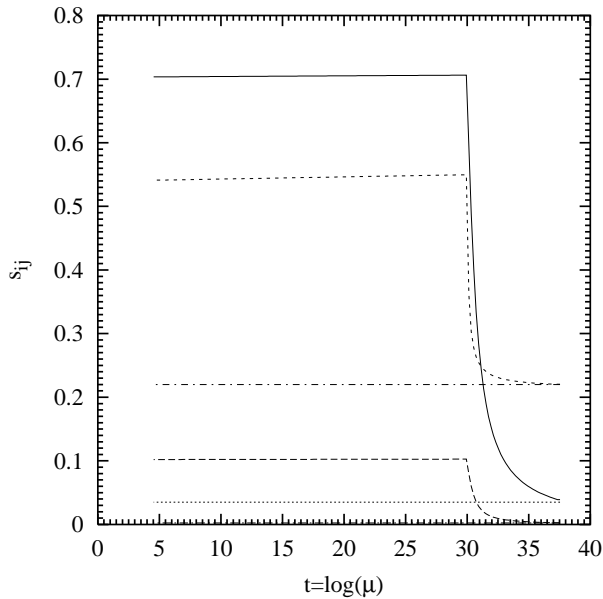}
\caption{Same as Fig.4 but with
$M_R=2\times 10^{16}$ GeV  and $M_{\rm S}$=$10^{13}$
GeV with input and output masses and mixing angles as given in the third column of Table 1.}
\label{fig5}
\end{figure}

\begin{acknowledgments}
M.K.P. thanks the Institute of Mathematical Sciences for Senior 
Associateship.
The work of R.N.M is supported by the NSF grant No.~PHY-0354401. The work
of G.R. is supported by the DAE-BRNS Senior Scientist Scheme of Govt. of
India.
\end{acknowledgments}


\begin{thebibliography}{99}
\bibitem{seesaw} P. Minkowski, Phys. Lett. {\bf B67 }, 421
(1977); M.~Gell-Mann, P.~Ramond, and R.~Slansky, \emph{Supergravity}
(P.~van Nieuwenhuizen et al. eds.), North Holland, Amsterdam, 1980,
p.~315; T.~Yanagida, in \emph{Proceedings of the Workshop on the
Unified Theory and the Baryon Number in the Universe} (O.~Sawada
and A.~Sugamoto, eds.), KEK, Tsukuba, Japan, 1979, p.~95; S.~L.
Glashow, \emph{The future of elementary particle physics}, in
  \emph{Proceedings of the 1979 Carg{\`e}se Summer Institute on Quarks and
  Leptons} (M.~L{\'e}vy et al. eds.), Plenum Press, New York, 1980,
pp.~687; R.~N. Mohapatra and G.~Senjanovi{\'c}, Phys. Rev. Lett. 
\textbf{44} 912 (1980).

\bibitem{so10} H. Georgi, {\it Particles and Fields}, ed. by C. E. Carlson
AIP, 1975); H. Fritzsch and P. Minkowski, Ann. Phys. {\bf 93}, 193 (1975). 

\bibitem{ps} J. C. Pati and A. Salam, Phys. Rev. {\bf D10}, 275 (1974).

\bibitem{ref4} K.S.~Babu, C.N.~Leung and J.~Pantaleone, Phys.~Lett.~{\bf
B319}, 191 (1993); P. Chankowski and Z. Pluciennik, Phys. Lett. {\bf B316},
312 (1993); S.~Antusch, M.~Drees, J.~Kersten, M.~Lindner and M.~Ratz,
  Phys.\ Lett.\ B {\bf 525}, 130 (2002) [arXiv:hep-ph/0110366];
S.~Antusch, J.~Kersten, M.~Lindner, M.~Ratz and M.~A.~Schmidt,
  JHEP {\bf 0503}, 024 (2005) [arXiv:hep-ph/0501272].

\bibitem{ref5} J.A.~Casas, J.R.~Espinosa, A.~Ibarra and I. Navarro,
~Nucl. Phys. {\bf B569}, 82 (2000); ~hep-ph/9910420.

\bibitem{mpr1} K.R.S.~Balaji, A.S.~Dighe, R.N.~Mohapatra and M.K.~Parida,
Phys.~Rev.~Lett.~{\bf 84}, 5034 (2000); Phys.~Lett.~{\bf B481}, 33 ~(2000);
K.R.S.~Balaji,  R.N.~Mohapatra, M.K.~Parida and ~E. A. Paschos,
Phys.~Rev.{\bf D63}, 113002~(2001).
\bibitem{mpr2} R. N. Mohapatra, M. K. Parida and G. Rajasekaran, hep-ph/
0301234; Phys. Rev. {\bf D69}, 053007 (2004).

\bibitem{seesaw2}  G. Lazarides, Q. Shafi and C. Wetterich,
Nucl.Phys.{\bf B181}, 287 (1981); R. N. Mohapatra and G. Senjanovi\'c,
Phys. Rev. {\bf D 23}, 165 (1981).
 

\bibitem{ref3}  M.~Appollonio {\it et~al.},~Phys.~Lett.~{\bf B466}, 415 
(1999);~F. Boehm ~et~al.,Phys. Rev. {\bf D64}, 112001 (2001).

\bibitem{bb} C.~Aalseth {\it et al.}, arXiv:hep-ph/0412300.
 
\bibitem{ref9} H.V. Klapdor-Kleingrothaus et al. Eur. Phys. J. {\bf A 12},
147 (2001); C. Aalseth et al. Phys. Rev. {\bf D 65}, 092007 (2002);
H. V. Klapdor-Kleingrothaus et al., Mod. Phys. Lett. {\bf
A 16}, 2409 (2001); hep-ph/0303217;
H.V. Klapdor-Kleingrothaus, A. Dietz and I.V. Krivosheina,
Foundations of Physics {\bf 32}, 1181 (2002).

\bibitem{ref10} A. Osipowicz et~al.,(KATRIN Project), hep-ex/0109033.

\bibitem{ref11} D. Spergel et al. Astro-ph/0302209; C. L. Bennett {\it et al.}
astro-ph/0306207; S. Hannestad, astro-ph/0303076;


\bibitem{ref12} R.~N.~Mohapatra, M.~K.~Parida and G.~Rajasekaran,
  Phys.\ Rev.\ D {\bf 71}, 057301 (2005) [arXiv:hep-ph/0501275].

\bibitem{ref6} D.~Sinclair  [SNO Collaboration],
  Nucl.\ Phys.\ Proc.\ Suppl.\  {\bf 137}, 15 (2004); A.~Suzuki  [KamLAND 
Collaboration],
  Nucl.\ Phys.\ Proc.\ Suppl.\  {\bf 137}, 21 (2004).

\bibitem{ref13} N. Arkani-Hamed and S.Dimopoulos,hep-ph/0405159;
hep-ph/0501082;

\bibitem{others} For some works on split supersymmetry, see G. F. Giudice 
and A. Romanino, hep-ph/0406088;
 N. Arkani-Hamed and S.Dimopoulos, G. F. Giudice and A. Romanino,
hep-ph/0409232; K.~S.~Babu, T.~Enkhbat and B.~Mukhopadhyaya,
  arXiv:hep-ph/0501079;  B.~Kors and P.~Nath,   Nucl.\ Phys.\ B {\bf 711}, 
112 (2005) [arXiv:hep-th/0411201]; M. Drees, hep-ph/0501106;
  A.~Pierce, Phys.\ Rev.\ D {\bf 70}, 075006 (2004)
  [arXiv:hep-ph/0406144]; A. Ibarra, hep-ph/0503160; R.~Mahbubani,
 arXiv:hep-ph/0408096.
\bibitem{ref16} V. Barger, M. S. Berger, and P. Ohmann, Phys. Rev. {\bf D47},
1093 (1993); C. R. Das and M. K.Parida, Eur. Phys. J. {\bf C 20}, 121
(2001).
\bibitem{ref17} D. G. Lee and R. N. Mohapatra, Phys. Lett. {\bf B 329},
463 (1994).

\bibitem{drees}  M. Drees, hep-ph/0501106; Katri Huitu, Jari Laamanen,
Probir Roy, Sourov Roy,  hep-ph/0502052.
.
\bibitem{ref18} E. J. Chun and S. Pokorski, Phys. Rev. {\bf D62},053001 (2000);
\bibitem{ref19}P.Chankowski and P. Wasowicz, Eur. Phys. J. {\bf C23}, 249
(2002).
\bibitem{PDG} S. Eidelman {\it et al.} Phys. Lett. {\bf B592}, 1 (2004). 

\bibitem{reactor} K.~Anderson {\it et al.},
arXiv:hep-ex/0402041; M.~Apollonio {\it et al.}, Eur.\ Phys.\ J.\
C {\bf 27}, 331 (2003) arXiv:hep-ex/0301017.

\bibitem{bnl} M.~V.~Diwan {\it et al.}, Phys.\ Rev.\ D {\bf 68}, 012002
(2003) arXiv:hep-ph/0303081; D. Ayrea et al. hep-ex/0210005;
Y. Itow et al. (T2K collaboration) hep-ex/0106019; I.~Ambats {\it et al.}
(NOVA Collaboration), FERMILAB-PROPOSAL-0929.

\end{thebibliography}
\end{document}